\documentclass{aastex631}
\newcommand{\gtsim}{\raisebox{-1.0ex}{$\stackrel{\textstyle>}{\sim}$}}
\newcommand{\ltsim}{\raisebox{-1.0ex}{$\stackrel{\textstyle<}{\sim}$}}
\long\def\comment#1{}
\def\kms{km~s$^{-1}$}

\def\hinode{{\sl Hinode}}

\def\p78{{\sl P78-1}}

\def\sdo{{\sl SDO}}

\def\psp{{\sl PSP}}

\def\feix{Fe~{\sc ix}}
\def\fexii{Fe~{\sc xii}}
\def\fexv{Fe~{\sc xv}}

\def\heii{He~{\sc ii}}

\def\kms{km~s$^{-1}$}

\begin{document}
%

\title{Inconspicuous Solar Polar Coronal X-ray Jets as the Source of Conspicuous \hinode/EUV Imaging Spectrometer 
(EIS) Doppler Outflows}

\author{Alphonse C. Sterling}
\affiliation{NASA/Marshall Space Flight Center, Huntsville, AL 35812, USA}

\author{Conrad Schwanitz} 
\affiliation{Institute for Particle Physics and Astrophysics, ETH Z{\"u}rich, 8092 Z{\"u}rich, Switzerland}
\affiliation{Physikalisch Meteorologisches Observatorium Davos, World Radiation Center, 7260 Davos, Switzerland}

\author{Louise K. Harra} 
\affiliation{Physikalisch Meteorologisches Observatorium Davos, World Radiation Center, 7260 Davos, Switzerland}
\affiliation{Institute for Particle Physics and Astrophysics, ETH Z{\"u}rich, 8092 Z{\"u}rich, Switzerland}

\author{Nour E. Raouafi} 
\affiliation{The John Hopkins University Applied Physics Laboratory, 
Laurel, MD 20723, USA}



\author{Navdeep K. Panesar} 
\affiliation{Bay Area Environmental Research Institute, NASA Research Park, Moffett Field, CA 94035, USA}
\affiliation{Lockheed Martin Solar and Astrophysics Laboratory, 3251 Hanover Street, Building 252, Palo Alto, CA 94304, USA}

\author{Ronald L. Moore} 
\affiliation{Center for Space Plasma and Aeronomic Research, \\
University of Alabama in Huntsville, Huntsville, AL 35805, USA}
\affiliation{NASA/Marshall Space Flight Center, Huntsville, AL 35812, USA}


\begin{abstract}

We examine in greater detail five events previously identified as being sources of strong transient coronal 
outflows in a solar polar region in \hinode/EUV Imaging Spectrometer (EIS) Doppler data.  
Although relatively compact or
faint and inconspicuous in \hinode/Soft X-ray Telescope (XRT) soft-X-ray (SXR) images and in 
Solar Dynamics Observatory (\sdo)/Atmospheric Imaging Assembly (AIA) EUV images, we find that all of 
these events are consistent with being faint coronal X-ray jets.  The evidence for this is that the events 
result from eruption of minifilaments of projected sizes spanning 5000---14{,}000\,km and with
erupting velocities spanning 19---46\,\kms, which
are in the range of values observed in cases of confirmed X-ray polar coronal hole jets.  In SXR images, and in 
some EUV images, all five events show base brightenings, and faint indications 
of a jet spire that (in four of five cases where determinable) moves away from the 
brightest base brightening; these properties are common
to more obvious X-ray jets.  For a comparatively low-latitude event, the minifilament erupts from 
near ($\ltsim$few arcsec) a location of near-eruption-time opposite-polarity magnetic-flux-patch convergence, 
which again is consistent with many observed coronal jets. Thus, although too faint to be identified as 
jets {\it a~priori}, otherwise all five events are identical to typical coronal jets.  This suggests that jets 
may be more numerous than recognized in previous studies, and might contribute substantially to solar wind outflow, and to the population of magnetic switchbacks observed in Parker Solar Probe (PSP) data.

\end{abstract}


\section{Introduction}
\label{sec-introduction}

Since the discovery of the solar wind it has been apparent that there is a nearly continuous
outflow from the Sun.  Recent observations from the Parker Solar Probe  (\psp)
\citep[e.g.][]{bale.et19} show that previously discovered magnetic disturbances in the solar wind
magnetic field called {\it switchbacks}  \citep[e.g.][]{yamauchi.et04} are extremely common. 
Exactly where the outflows, and features such as switchbacks, originate and how they are generated 
are outstanding puzzles.

Movies constructed from high-cadence images in lines sensitive to 
coronal emissions reveal a candidate source of those outflows and/or switchbacks,
namely the transient outbursts known as {\it coronal jets}.  These features take the form of
long and narrow columns of emission  at soft X-rays (SXR) and/or EUV wavelengths
\citep[][]{shibata.et92,cirtain.et07}.  Observing in SXRs with the
X-Ray Telescope (XRT) on the \hinode\ spacecraft,  \citet{savcheva.et07} report
that there are $\sim$60 jets/day in  the two polar coronal holes.  Assuming that jets are
distributed evenly over the entire solar  surface, taking the coronal holes
to cover roughly 15\% percent of the solar surface 
\citep{sheeley80,harvey.et02}, and extrapolating to the entire Sun, gives only several 
hundred jets per hour, which is far from enough to
explain the large quantities of switchbacks observed by PSP \citep[e.g.,][]{bale.et19}.  
If jets are a major source for switchbacks, and also perhaps of solar wind material, 
then they must be more frequent than the \citet{savcheva.et07} values.

\comment{

Jets on smaller size scales,
called {\it jetlets} \citep{raouafi.et14,panesar.et18b}, could be the source for
more switchbacks.  If some spicules are small-scale versions of the same
jets, then they could increase the potential number of jet-produced switchbacks
even further \citep{sterling.et20b,neugebauer.et21}.

}

Coronal outflows can also manifest in EUV spectral data, as blue shifts in 
spectral lines.  Recently,  \citet{schwanitz.et21} used data from the 
EUV Imaging Spectrometer (EIS) instrument on the \hinode\ spacecraft to
look for such outflows.  They found 14 outflow events in the data sets they
examined (\S\ref{sec-data} provides more details on their data set).  

Using this data set of EIS blueshifted events, \citet{schwanitz.et21} then
performed an initial investigation of coronal EUV images, and also in SXR images 
for some events, to identify the source of the EIS blueshifts in those images.  Among the entire
set of 14 blueshifted events, they identified only five (events~1, 6, 11, 12, and 13 in their list) 
as corresponding 
to ``obvious jets or bright points" in those images. And specifically, they 
identified only one of the 14, event~1 in their list \citep[Table~1 in][]{schwanitz.et21}, 
as coinciding with an ``obvious jet." This, therefore, implies that there might be
frequent strong localized coronal outflows that occur in the low corona that are
not due to features clearly identifiable as jets in the coronal images.  This seemingly introduces 
a new puzzle, as to the cause of outflows if they are not jets.

In this paper, we will inspect more closely in coronal images a subset of the EIS blueshifted
events of \citet{schwanitz.et21}.  Our subset consists of the first five of
those events, labeled as events~1---5 in their paper.  We will examine imaging
data from SXRs and several EUV channels to see whether the source of these EIS
blueshifted events can be identified with jets or jet-like features, when
examined more extensively than in the \citet{schwanitz.et21} initial
investigation.  We limit our investigation to just the first five of their events 
because they are the only ones in the  data set of 14 that include SXR data. By including the SXR
data, we will be able to compare any possible jet signatures that we may find
from the EIS blueshifted events more directly with earlier SXR-jet
studies  \citep[e.g.,][]{shibata.et92,shimojo.et96,cirtain.et07,
savcheva.et07,sterling.et15,sterling.et22}.  Also, jets that appear obvious in SXRs do not always
manifest as obvious jet features in all EUV channels
\citep{moore.et10,moore.et13,sterling.et15,sterling.et22}, and so by including both SXR and EUV
images we will have a better chance of finding a jet corresponding to any of the five EIS Doppler 
events, if such a jet exists.

Central to our investigations will be our current understanding of how jets form.  As noted
above, the first detailed jet studies were with SXR observations. Among other findings, these
observations revealed that jets have both a long spire that extends away from the surface, and
a  base region that is bright in SXRs.  For many jets, the brightest portion of the bright base
is localized in the base on one side of the spire \citep[e.g.,][]{shibata.et92}.  Here we shall
refer to this brightening as a ``jet bright point" (JBP), following \citet{sterling.et15}.

Observations in the EUV provided insight into the workings of many jets that are not readily 
apparent in SXR images alone.  These wavelengths, at sufficient spatial and temporal resolution,
show that many coronal jets result from eruptions of miniature filaments, or minifilaments
\citep[e.g.,][]{nistico.et09,moore.et10,moore.et13,shen.et12,adams.et14}.
Based on observations of 20 jets in polar coronal holes, 
\citet{sterling.et15} argued that essentially all coronal jets result
from minifilament eruptions, and they further argued that the JBP was actually 
a miniature flare, that forms below the erupting minifilament in a manner analogous to 
flares that occur below erupting large-scale filaments in typical solar eruptions.  The 
jet-producing erupting minifilaments are essentially invisible in SXRs, which is why their 
importance was not recognized earlier.

Other studies have shown that the minifilaments erupt from miniature magnetic
neutral lines, frequently where magnetic flux cancelation is occurring 
\citep[e.g.,][]{shen.et12,hong.et14,young.et14a,young.et14b,adams.et14}.  A
series of papers
\citep{panesar.et16a,panesar.et17,panesar.et18a,mcglasson.et19}  shows that flux
cancelation frequently accompanies jet production.

The schematic picture of \citet{sterling.et15} has been presented several times,
and so it will not be repeated here (Fig.~1 of 
\citeauthor{sterling.et18}~\citeyear{sterling.et18} gives a
slightly updated version of the \citeauthor{sterling.et15}~\citeyear{sterling.et15}
schematic). Here we will describe the process assuming that the jet is occurring
in a polar coronal hole's open-field.  We use this assumption because (i) all of 
the jets of this paper occur in or around a polar coronal hole, and (ii) because assuming an open field
simplifies the description somewhat.  Actually however, the same basic process can 
hold for quiet Sun and active regions, but in those cases the field may be closed 
but far reaching instead of open.  Jets are, in fact, also common in quiet Sun and 
even at the edge of active regions, and they likely operate in the fashion that we now describe.

Basically, the \citet{sterling.et15} picture says that polar coronal jets occur in
regions of open magnetic field, where that field is rooted in flux of 
the hole's dominant magnetic polarity; for this discussion we will have the open-field
polarity be negative.  The picture assumes \citep[based on on-disk observations of 
jets, such as that of][]{adams.et14} that an opposite-polarity (positive) magnetic flux 
element sits in
this sea of negative flux. An ``anemone" type of magnetic structure \citep{shibata.et07} emanates
from the positive patch in small loops from the positive element to the surrounding negative 
flux.  This anemone structure is encased in negative-polarity open field, with the resulting 
structure having a magnetic 
null between the closed-field anemone and the open field above it. One side --- appearing as a lobe in a 
2-D cross-sectional cut --- of the 
anemone develops a minifilament/flux rope on the neutral line between positive- and 
negative-polarity patches, while the other (larger-lobe) side of the anemone is nearly potential
and does not have a minifilament. That minifilament/flux rope and its enveloping lobe field 
then erupt and move up along the outside 
of the potential-field lobe, until the erupting lobe field runs into oppositely 
directed open field on the far side of the potential-field lobe.

Two reconnections ensue during this minifilament eruption process: One is when
the erupting lobe field enveloping the erupting minifilament/flux rope encounters
the open field above the potential-field anemone lobe.  This reconnection produces outflow of heated
coronal plasma along the open field, forming a hot jet spire that is detected
in SXRs and in some of the hotter EUV channels.  \citet{sterling.et15} 
call this ``external reconnection" (aka ``interchange" or ``breakout" reconnection), 
because it occurs external to the lobe field
enveloping the erupting minifilament.  A second reconnection, called ``internal
reconnection" occurs between
the imploding legs of that erupting enveloping
field.  \citet{sterling.et15} argue that this internal reconnection results 
in the JBP, which corresponds to a typical solar flare that forms beneath an erupting 
typical-sized filament.  

In followup studies to the \citet{sterling.et15} work, \citet{panesar.et16a} and 
\citet{panesar.et17} argue that this minifilament/flux rope production and eruption result from 
magnetic flux cancelation along the neutral line, and that such cancelation first results in the formation of 
the minifilament, and that further cancelation leads to destabilization and eventual eruption of
the minifilament to make the jet. 

At least many jets are consistent with this ``minifilament eruption" model, and with
flux cancelation being responsible for the production and eruption of the minifilament/flux rope.  
For example,
\citet{mcglasson.et19} examined 60 on-disk coronal hole and quiet Sun  jets, and
found 85\% of them resulted from  minifilament
eruptions that coincided closely in time to underlying  flux cancelation.  One
study \citep{kumar.et19} argues that actual  flux cancelation is important only
in a small percentage ($\sim$20\%) of jets, with pre-jet field stressing by shearing and/or rotational
photospheric motions playing a more important role.  Another recent study however
\citep{muglach21} also maintains that actual cancelation does coincide with most
jets in that study. All of these studies are consistent with the
minifilament-eruption picture and flux cancelation, or at least flux
convergence and shearing, occurring for many jets in coronal holes and quiet
Sun.  Jets occurring at the periphery of active regions
\citep[e.g.,][]{mulay.et16,sterling.et16b} may operate in the same way as 
described above for coronal hole and quiet Sun jets, but with somewhat different
observational manifestations  due to the stronger and more dynamic fields at
those locations \citep{sterling.et17}. Numerical simulations recreate many of 
the aspects of this minifilament-eruption picture for producing jets
\citep{wyper.et17,wyper.et18a}; those simulations emphasize the breakout
reconnection (which we call the external reconnection above) in the formation of
the jets, but the fundamental processes involved are as in the minifilament
eruption picture of \citet{sterling.et15}.  Several reviews of coronal jets are
available
\citep[e.g.,][]{shimojo.et00,shibata.et11,raouafi.et16,hinode.et19,innes.et16,sterling18,shen21,sterling21,schmieder22}.

In the following, we will examine EIS Doppler events~1 through 5 of
\citet{schwanitz.et21}.  We will argue that the evidence indicates that each of
these events originate from hard-to-detect jets that fit the minifilament-eruption 
picture for jets described above.

\section{Instrumentation and Data}
\label{sec-data}

In this study we will use SXR images from \hinode/XRT \citep{golub.et07}, and EUV images
from the Solar Dynamics Observatory (\sdo)'s Atmospheric Imaging
Assembly \citep[AIA;][]{lemen.et12}.  XRT is a grazing incidence X-ray telescope
that observes the Sun using several broad-band filters. It has a detector 
of $512\times 512$~square pixels, each $1\arcsec\kern-0.5em.\,02$ wide. Polar coronal jets, such 
as the features of relevance here, have been effectively viewed with XRT's
filters sensitive to temperatures of $\gtsim$1\,MK \citep{cirtain.et07}, and this
is fully consistent with measured temperatures of jet spires having values clustered
in the 1.5---2.0\,MK range \citep{pucci.et13,paraschiv.et15}.  Here we use
data obtained with XRT's ``Al-Poly" filter, from a run of the \hinode\ {\it operations
program} (HOP)~81.  This HOP has been run on a regular basis since near the time of
\hinode's launch in 2006, and focuses observations on the solar polar regions.
The resulting images in our data set have an approximately 30\,s cadence.  XRT
supports a variety of options for the field of view (FOV) of the images, and for these runs 
it was set to $384'' \times 384''$.  For most of our analysis, and for most of the XRT images
and movies presented here, we use cutouts of this full area to examine the features
of interest more closely.

\sdo/AIA routinely observes the entire solar disk in seven EUV filters.  Here we examine in
detail images from four filters, centered on wavelengths of \heii\ 304,  \feix\ 171, \fexii\
193, and \fexv\ 211~\AA\@.  In contrast to the XRT Al-Poly filter, which is broadband and responsive
to essentially all coronal temperatures in excess of $\sim$1\,MK, AIA filters have a narrower 
temperature response; the 304, 171, 193, and 211\,\AA\ AIA filters to be used here are most sensitive  to non-flaring
coronal temperatures of approximately 50,000\,K, $6\times 10^5$\,K, $1.5\times 10^6$\,K, and 
$2.0\times 10^6$\,K, respectively.  We restrict ourselves to these four AIA filters because
\citet{sterling.et15} found that the hotter-response AIA EUV channels (131, 335, and 94\,\AA)
added little to the information obtainable from our four selected channels for the type of
investigations to be done here in polar coronal hole regions.

We will be looking for faint features in the images, and we found it helpful to enhance
them by performing a running sum over two successive images, while retaining the 
native cadences for the animations of the XRT and AIA images.  We also enhance the
contrast of weak features by taking the fourth root of the image intensities, before
displaying them with the usual logarithmic scaling.  We adjust the maximum and minimum
intensity levels as needed for each event we examine to highlight features of interest.
This same display technique was used in \citet{sterling.et22}.

\begin{figure}
\centering
\includegraphics[width=\textwidth]{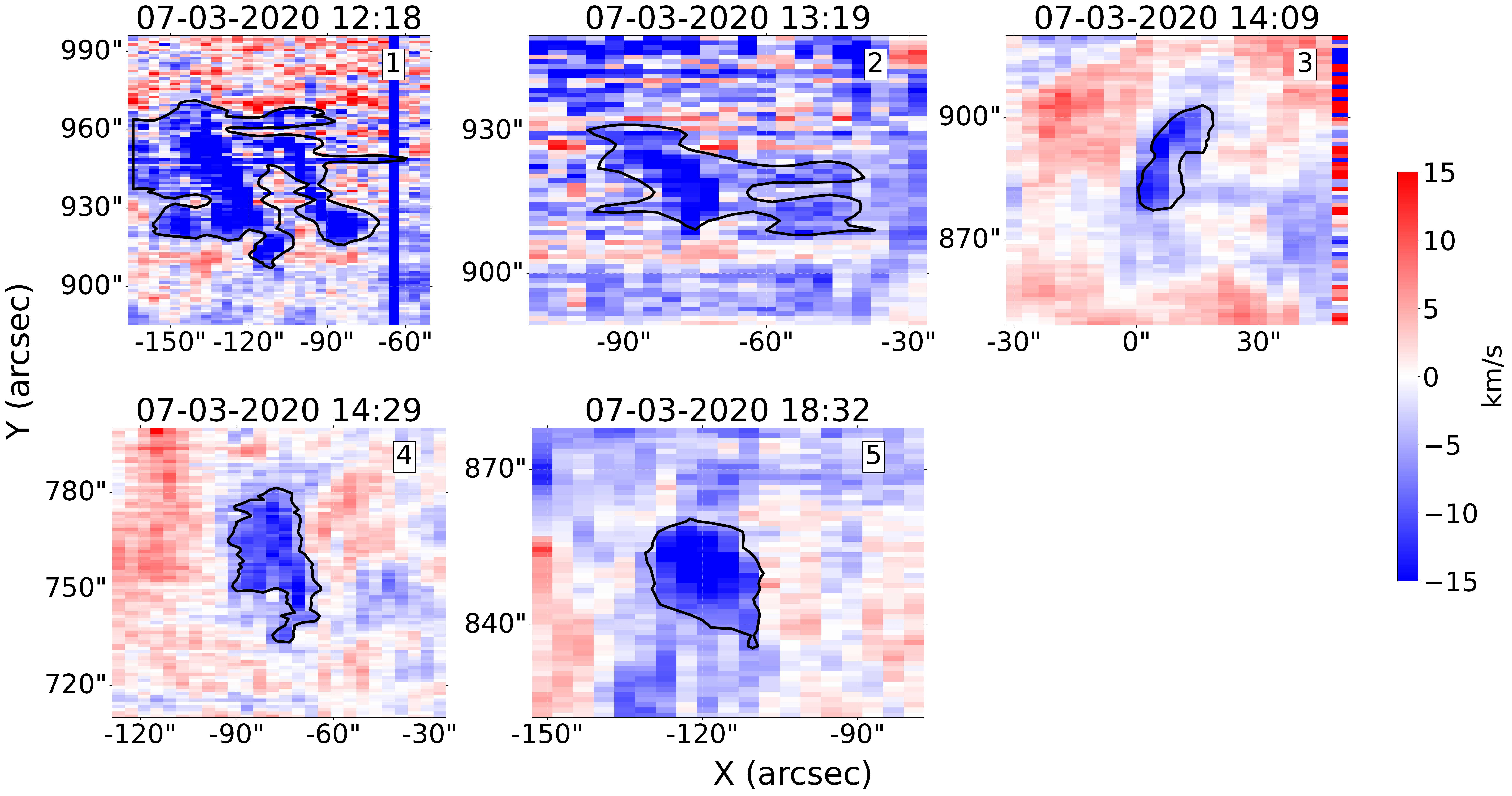}
\caption{EIS-observed Doppler outflow events for the five events studied here, in the \fexii~195.12\,\AA\ 
emission line.  These are the first
five events studied in \citet{schwanitz.et21}, with the event number (which also corresponds to the
event numbers of this paper) given in the upper right of each panel, and with each event's date
and time of the center of the Doppler outflow observations given at the top of each panel in 
DD-MM-YYYY HH:MM format.  Deep blue
features inside the contours represent observed outflows in excess of Doppler velocity $-6$\,\kms.
See \S\ref{sec-data} and \citet{schwanitz.et21} for other details.}
\label{eis_overview}  
\end{figure}

The events were originally selected by \citet{schwanitz.et21} from 
\hinode/EIS studies from three observation 
periods, consisting of 12 raster scans in total.  
Here we will look at the features from the first of these observation periods, during which this
EIS study was also part of HOP~81.    EIS took those scans
on 2020 March 07, using the EIS $2''$ slit and exposure times of
50~s, consisting of five rasters of duration
70\,minutes each.  Spectra were obtained in the \fexii\ emission
line at 195.12\,\AA, and were processed and fitted with single Gaussian profiles
to obtain intensities and other information.  Based on a rest wavelength, 
determined by using the average centroid of the entire raster, 
\citet{schwanitz.et21} marked
times and locations that displayed blue shifts in excess of $-6$\,\kms.  In this
fashion, they identified five EIS blue-shifted events in the coronal hole data over the chosen period, which are shown in Figure~\ref{eis_overview}.
(There were 14 such identifications in the full set of scans
examined by \citeauthor{schwanitz.et21}~\citeyear{schwanitz.et21}.)
The so-identified blue-shifted events were transient, in the sense that
they did not exceed the duration of a single raster scan of 70~min, which
was the finest time cadence available in that set of EIS data.   
Additionally, the blueshifted locations were localized; the largest blueshifted
region among the five of the first scan was 3244 arcsec$^2$, and the other
four at 721 arcsec$^2$ or less.  Table~1 lists these five events, with the EIS time
of observation (the time of the center of the blueshift contour) in column~1 and the blueshifted areas in column~2.

Figure~\ref{xrt_overview_zu} shows an XRT image of the north polar coronal hole region during
the period of observation, at the time when event~3 is occurring.  Also overplotted is a box showing the approximate FOV of the panel for this event in Figure~\ref{eis_overview}.  The accompanying animation
shows the XRT observation over the entire period considered, with all five events examined here both labeled and including a box showing the FOV of the corresponding EIS panel in Figure~\ref{eis_overview}.  Table~1 gives in columns~3 and~4 the times and 
locations of the five events that we will inspect here in XRT and AIA images.

\section{Results}
\label{sec-results}

We consider each of the five EIS events, examining their XRT and AIA images in detail.
We found all five of these events to be consistent with originating in the same fashion
as typical coronal jets, namely via minifilament eruptions.  We found, however, the arguments for
this for cases events~3, 4, and~5 to be more straightforward than events~1 and~2.  Therefore we
will present our findings in that order.

\subsection{Event~3}
\label{subsec-results_e3}

Event~3's probable source was described by \citet{schwanitz.et21} (in that paper's Table~1) as a ``small-scale brightening."  We will see that there is indeed a compact brightening that accompanies this outflow event.  In the following, we argue that this is the brightening at the base of a faint, but otherwise typical, coronal jet.

Figure~\ref{event_03_zu} shows our results for event~3.  This is two six-panel arrays, with 
panels~(a)---(f) showing
the region around the event in different wavelength channels all at nearly the same time (within a few seconds), 
and panels~(g)---(l) showing the same FOV of the same channels all at nearly the same 
time (again within a few seconds), but where the (g)-(l) times are about one minute later than the (a)---(f) panels.  Each of the two sets shows images respectively in AIA~171, 193, 211, and~304\,\AA\ in panels (a),
(b), (d), and~(e).  Panels~(c) and~(f) show XRT images at the same time, with
the image in panel~(f) more zoomed-in than that in panel~(c).  Neither of the two XRT panels 
are as zoomed-in as the
AIA panels, because we could not effectively zoom in as close with XRT due to its lower spatial 
resolution than AIA\@.  An animation accompanying the on-line version shows the dynamic evolution 
of these panels over the course of event~3.

Figures~\ref{event_03_zu}(c) and~\ref{event_03_zu}(i) show XRT images close in time 
to Figure~\ref{xrt_overview_zu}, so that the 
brightening pointed to by the blue arrows in the Figure~\ref{event_03_zu} panels is identifiable 
in Figure~\ref{xrt_overview_zu}, just above the ``E3" label. In that larger FOV image of  
Figure~\ref{xrt_overview_zu} the feature seems nondescript.  In Figure~\ref{event_03_zu} however, we
have zoomed in enough to see that there is an obvious eruption occurring at that location.
Green arrows in Figures~\ref{event_03_zu}(a) and~\ref{event_03_zu}(g) point to this 
feature in 171\,\AA\ images,
showing clear evolution over this one-minute time difference.  The dynamic motion of this erupting
feature is obvious in the accompanying animation, where the motion becomes apparent in 171\,\AA\ at 
14:06:33\,UT, and continues until the end of the animation at 14:34:45\,UT\@.  In the animation, the
dynamic motion is apparent in the other three AIA channels as well.   

We have measured the size of this erupting feature, using the 304 and 211\,\AA\ channel images at
four different times between 14:08:21 and 14:09:05\,UT, obtaining a size of 
$7\arcsec\kern-0.5em.\,8 \pm 0\arcsec\kern-0.5em.\,95$ 
for the mean and standard deviation of those four measurements.
We also measured the speed of movement of the feature, using the same AIA channels and manually tracking
different parts of the feature, over approximately 14:07:29 and 14:09:05\,UT\@, giving 
$45.8 \pm 15.4$\,\kms\ for the mean and standard deviation of those four measurements.  All of these
measurements are of the proper motion of the feature in projection against the observed solar disk.

The observed erupting feature in the AIA channels, as well as its lack of visibility in soft X-rays,
are all consistent with the appearance of erupting minifilaments that produce coronal jets.  Moreover,
the measured length is within the range of the size measured
for erupting minifilaments that made features that are unambiguously defined as polar coronal hole jets, albeit near the lower end of that range.
\citet{sterling.et15} found for 18 such jets (out of their total set of 20 jets, with two of the events being too diffuse for adequate measurement) a size of $11'' \pm 7''$, and \citet{sterling.et22} found for
an independent set of 20 soft X-ray polar coronal hole jets (out of 21 total jets) that the erupting 
minifilaments had size $14'' \pm 7''$. All of those measurements were made in a fashion similar to 
the method we use here; the specific AIA channel used for those measurements varied, but was among the same four
channels used here.  We 
believe that the differing channels is not a factor in the measurement in these cases however, because we selected
a channel or channels where the cool-material erupting feature could be well seen.  The AIA channel 
of best erupting-minifilament visibility varies from event to event, as discussed in some detail 
in \citet{sterling.et22}.

Similarly, the speed that we find for for the erupting absorbing feature of 
event~3, $\sim$45\,\kms, is similar to those that we
measured for erupting minifilaments at the base of X-ray polar coronal hole jets in our previous studies: the 18 
such jets of \citet{sterling.et15} had velocities of $31 \pm 15$\,\kms; and while the 20 jets of 
\citet{sterling.et22} had average velocities and standard deviation of $26 \pm 13$\,\kms, two of those jets had values in excess of
45\,\kms.  We have used the same general measurement techniques for the cool-material erupting features that
we see here as we did in those two previous studies of the erupting minifilaments that we found
at the base of the polar coronal hole jets.  Because of the similarity in appearance, size, and 
velocity of the erupting feature causing event~3 to erupting minifilaments that we have identified previously, 
henceforth we will regard the dark feature that we see erupting in event~3 as an erupting minifilament.

Moreover, the brightening pointed to by the blue arrows in the Figure~\ref{event_03_zu} XRT 
panels~(c) and~(i),
and also visible as brightenings in panels of the AIA wavelengths, is similar to brightenings that we
see at the base of erupting minifilaments.  Comparison between the panels among the different 
wavelengths of the animation accompanying 
the figure also show that the erupting minifilament was originally seated near the solar surface at the
location where the brightening later starts (at about 14:00\,UT) in soft X-rays.  

Because properties of the apparent erupting minifilament of event~3 are similar to erupting minifilaments that make confirmed jets, this similarity motivates the question of whether the event~3 erupting minifilament generated a coronal jet.  The images and
animation of Figure~\ref{event_03_zu} argue that it does.  White arrows in panels~(h) and~(i) show
a feature that appears east of the location of the erupting minifilament. This feature is only weakly 
visible in the XRT images, but its structure persists for several frames in the animation.  Similarly,
the feature shows up well in the 193\,\AA\ AIA images.  This feature is fully consistent with being 
the spire of a coronal jet. Moreover, the brightening pointed out by the blue arrows is then 
consistent with being a JBP typically seen accompanying coronal jets. In addition, the apparent 
spire moves with time over its life, consistent with what has been observed in jets 
\citep{shibata.et92,savcheva.et07}, and moreover, that movement with time is away from the 
location of the suspected JBP \citep{savcheva.et07}, and this is consistent with a jet resulting from
the erupting minifilament \citep{baikie.et22}. 

Figure~\ref{event_03_dim_zu}, and its corresponding animation, show fixed-frame difference images of
this event.  Figure~\ref{event_03_dim_zu}(a) reveals a feature that looks the same as a spire 
of typical coronal jets,
and the animation shows that it moves away from the JBP just as in typical coronal jets.  Therefore
we conclude that our event~3 is indeed a faint version of a coronal jet.  

Figure~\ref{event_03_dim_zu}(b) shows
a frame toward the end of the corresponding animation, with a region at the base showing a clear
dimming signature.  Such dimmings at coronal wavelengths can be indicative of showing locations where
previously closed magnetic fields become temporarily open \citep[e.g.,][]{sterling.hudson97,zarro.et99}.
Thus this dimming is consistent with material previously trapped near the solar surface escaping
out along open field lines, which could account for the dimming signatures observed in EIS\@.  
From the non-differenced animation accompanying Figure~\ref{event_03_zu}, the erupting minifilament
passes near the location of the dimming over about 14:08---14:10\,UT, while dimming in 
the Figure~\ref{event_03_dim_zu} video is over primarily 14:10---14:20\,UT; this supports that our observed dimming results from a field opening, rather than from a temporary obscuration of lower-height hot coronal material by cooler minifilament material passing above it.  Thus our observed
dimming accompanying our suspected jet of event~3 is a good candidate for being the source of the EIS outflows.

\subsection{Event~4}
\label{subsec-results_e4}

Event~4's probable source was described by \citet{schwanitz.et21} (in that paper's Table~1) as a ``small-scale eruption."  Our work below will substantiate that assessment.  We will further however argue that this small-scale eruption produces a weak (i.e., faint and comparatively hard-to-see) coronal jet.

Figure~\ref{event_04_zu} shows our results for event~4.  This again is two six-panel arrays, with 
the same layout as Figure~\ref{event_03_zu}.  An animation accompanying the on-line version 
shows the dynamic evolution of these panels over the course of event~4.

We have again selected the AIA FOV to highlight a feature we observe near the time of event~4.  In all four
displayed AIA channels, this feature clearly appears to be an erupting minifilament.  Arrows in panels~(a)
and~(e) show it respectively in 171 and 304\,\AA\ images, and panels~(g) and~(k) show it
having moved outward about four minutes later.  As with event~3, we measure the mean length and speed
of this feature, and find it to have size $17\arcsec\kern-0.5em.\,3 \pm 1\arcsec\kern-0.5em.\,5$, 
measured at three times in 304\,\AA\ 
images over the interval 14:09:53---14:15:17\,UT\@.  We find a mean speed projected against the
solar disk from 304\,\AA\ images over 14:11:29---14:12:41\,UT to be $29 \pm 10$~\kms.  As with
event~3, these values for the erupting cool material feature are within the ranges previously 
measured for X-ray-coronal-jet-producing minifilament eruptions.  Thus, all indications are that 
this event-4 feature is equivalent to such an eruption.

Figures~\ref{event_04_zu}(i) and~\ref{event_04_zu}(l) show XRT frames with a very faint feature (white arrows), 
that appears to be consistent with a weak jet spire.  Blue arrows in the same panels show a 
feature consistent with being a JBP accompanying the possible spire.  From the animation 
accompanying that figure, that brightening is transient, appearing over about 14:13---14:29\,UT\@.
It occurs roughly in tandem with the prospective spire, and the prospective spire moves away
from the brightening with time, over about 14:20---14:28\,UT\@.  All of this is consistent with 
the prospective spire again being
a faint coronal jet spire, and the brightening being a typical JBP commonly occurring at the
base of coronal jets. Returning to the AIA images, weakly enhanced intensity outflows consistent
with this spire are discernible in 193 and 211\,\AA\ videos accompanying Figure~\ref{event_04_zu},
with the white arrows in Figure~\ref{event_04_zu}(h) pointing to the feature.  From the videos,
the outward movement of this feature has a large component of motion toward Earth.  This likely 
is an additional factor in the extreme weakness of the spire in the images and videos.  These observations support that this feature is a faint version of a coronal jet instigated by an erupting 
minifilament.

Clear apparent outflows are visible in the Figure~\ref{event_04_zu} AIA~193\,\AA\ video 
over about 14:13---14:30\,UT, leading to an intensity reduction (dimming) between the location 
of the brightening and the jet-spire feature pointed to by the arrows in Figure~\ref{event_04_zu}(h).
Again, this dimming is consistent with being the source of the outflows recorded in the EIS Doppler spectral data.

These factors all support that event~4 is a faint coronal jet, and that it was the source of 
the corresponding EIS Doppler outflows.

\subsection{Event~5}
\label{subsec-results_e5}

Event~5's probable source was described by \citet{schwanitz.et21} (in that paper's Table~1) as ``unclear."  We will argue here that the outflow is due to outflows along the spire of a coronal jet.

Figure~\ref{event_05_zu} shows our results for event~5, with 
the same layout as Figures~\ref{event_03_zu} and~\ref{event_04_zu}.  An animation 
accompanying the on-line version 
shows the dynamic evolution of these panels over the course of event~5.  

With the zoomed-in
AIA images and animations, we again see clear evidence for an erupting minifilament in all
four AIA channels.  It might be best seen in 304\,\AA\ over 17:55---17:59\,UT, crossing 
a bright feature.  More generally, it is apparent in the 304\,\AA\ video over
17:53---18:00\,UT\@. It is visible in the other AIA channels at the same time but with somewhat
less contrast (with the scalings used here).  We have measured the feature's length in 304\,\AA\ at 
three times over 17:56----17:59\,UT, obtaining $19\arcsec\kern-0.5em.\,3 \pm 0\arcsec\kern-0.5em.\,58$.  
For its speed projected
against the disk, we used 304\,\AA\ images for three measurements over 17:55:05---17:57:53\,UT, 
obtaining $32 \pm 11$~\kms.  Again, these values are consistent with the feature being an 
erupting minifilament.

White arrows in Figure~\ref{event_05_zu}(l) point to either side of a spire-like feature.  In
the accompanying XRT video, there is clear evidence for a spire at that arrowed location starting
shortly after the erupting minifilament occurs, expanding outward over 18:17---18:31\,UT\@.
Blue arrows in Figures~\ref{event_05_zu}(i) and~\ref{event_05_zu}(l) show a brightening with 
a burst of intensity
over about 18:15---18:20\,UT, in tandem with what looks like an opening up and expansion of
the spire; this is consistent with being a jet-base JBP, with the jet spire again moving 
away from the JBP location (expanding away from the JBP over about 18:09---18:40\,UT\@).  
A clear dimming signature is apparent in the XRT video over 
about 18:22---18:41\,UT, occurring between the JBP and the spire location (the dark channel extending
from the JBP at an angle to the north and slightly to the east at 18:41\,UT was not present 
at 18:22\,UT)\@.  

In this case, the apparent spire is rooted about 30$''$ to the northeast of the brightening
(Fig.~\ref{event_05_zu}(l).  This is farther than the comparable distance in E3 (Fig.~\ref{event_03_zu}(i), where the separation is only a few arcsec), and in E4 (Fig.~\ref{event_04_zu}, where the separation is $\sim$5$''$). Inspection of the AIA
videos, however, shows that in this case there is a horizontal outflow that hugs the
solar surface that originates near the minifilament-eruption location (where the JBP forms), out to near where the spire occurs.  We show this movement with an arrow in the 171\,\AA\ video corresponding to Figure~\ref{event_05_zu}.  The JBP in this case is well outside of the EIS box.  One portion of that box is visible in the zoomed images of Figure~\ref{event_05_zu}, but we represent part of that box in frames (k) and (l) as blue lines; see the video accompanying Figure~\ref{xrt_overview_zu}, which also includes those blue lines in the the XRT panel (during the first run of the video).  Figure~\ref{event_05_zu}(l) shows that the spire occurs inside of the EIS outflow box, even though the JBP is outside of that box.  \citet{sterling.et22}, in Fig.~5 of that paper along with the accompanying animation, shows another example of substantial horizontal movement
of an erupting minifilament, before the eruption turns outward to a spire some distance 
away from the JBP location. 

\comment{
In this event~5 case, we find the spire to be more obvious in the XRT video presented here. We have seen before
cases where the spires of  XRT jets in polar regions are not particularly prominent in AIA
videos \citep{sterling.et22}. 
} 

It often is not possible to obtain useful line-of-sight HMI magnetic field information from 
polar-region features, because the photospheric fields detected by HMI are oriented primarily 
in the radial direction, meaning that the Earth-directed line-of-sight component is small with
most of the true signal  lost in the noise.  We have found however that event~5 is at a low-enough
latitude and that the region contains strong-enough fields for such magnetograms to be of qualitative
use.  Figure~\ref{event_05_hmi_zu} shows these HMI magnetograms.  
Figure~\ref{event_05_hmi_zu}(a) shows contours from about the time of 
Figure~\ref{event_05_hmi_zu}(d), overlaid onto a grayscale version of a further-zoomed-in portion of the 304\,\AA\ image of 
Figure~\ref{event_05_zu}(e).   There are two relatively strong opposite-polarity 
magnetic-flux patches very near the erupting-minifilament location.  Reviewing the video 
accompanying Figure~\ref{event_05_zu}, it can be seen that the erupting minifilament erupts
from a location near the neutral line of those flux patches, starting sometime before when the spire
becomes discernible in the 304\,\AA\ images at about 17:54\,UT\@.  At that time it is about 
$3''$ away from that neutral line, and that is likely comparable to the alignment-offset 
uncertainty between the magnetograms and the AIA images.  We can only say that the erupting 
minifilament originates from, and the JBP occurs at, a location close to the neutral line of 
that bipolar strong-flux patch.

Figure~\ref{event_05_hmi_zu}(b---d) show the evolution in time of this magnetic region, showing that the
two flux patches approach each other in the hours prior to the minifilament eruption.  As pointed 
out in \S\ref{sec-introduction}, flux convergence and cancelation has been found to accompany jet
onset closely in time in several 
studies of on-disk jets in quiet-Sun and coronal holes 
\citep{panesar.et16a,panesar.et17,panesar.et18a,mcglasson.et19,muglach21}. And we have also observed
similar flux cancelation at the time and base location of active-region jets \citep{sterling.et16b,sterling.et17}.  In the present case, we 
see the patches approaching each other, and it is hard to confirm whether actual cancelation is
occurring. Frequently such cancelation that apparently triggers the minifilament's eruption 
appears to be among ``weak flux grains" near the neutral line of a larger-flux bipolar pair 
\citep{panesar.et18a}.  If such cancelation is occurring in our event~5 case here, it would
be beyond detectability in the line-of-sight magnetograms observing near the polar region.
Therefore, we can only say that the convergence of the opposite-polarity bipolar pair that 
we observe here is consistent with flux cancelation triggering the eruption of the minifilament,
as has been previously reported for more-on-disk coronal jets.  Nonetheless, the behavior of the
magnetic flux elements at the base of event~5 provides further evidence that it is a typical 
coronal jet, that is more faint than ones observed previously.

\subsection{Events~1 and~2}
\label{subsec-results_e1_e2}

We now consider events~1 and~2.  These two EIS detections occurred at nearly the same
location on the Sun, separated by about one hour in time.  The probable sources were described
as ``obvious jets" for event~1 and as a ``small-scale eruption" for event~2 by \citet{schwanitz.et21}.
Here we examine the features more closely.

From the video accompanying Figure~\ref{xrt_overview_zu}, the two events appear to be in
the same magnetically active location.  In Figure~\ref{event_01_02_zu} we show 
zoomed-in images, and accompanying videos, of the region encompassing both events.

\subsubsection{Event~1}
\label{subsubsec-results_e1}

Figures~\ref{event_01_02_zu}(a---f) show event~1. We observe a jet-like 
spire in the XRT images, pointed to by the white arrow in  Figure~\ref{event_01_02_zu}(f).  
From the accompanying video,
the 304\,\AA\ images show an erupting cool-material feature.  In this case, the feature 
is small and difficult to disentangle from the surrounding material.  It still however
looks similar to the erupting minifilaments that we detected in events~3---5.  It is
likely that this feature is less obvious than the erupting minifilaments of those previously
examined events in part due to its proximity to the limb in this case, combined with its
orientation, where we are viewing it looking nearly along its long axis.  That is, it 
resembles the erupting minifilament of, say, event~4 viewed in
304\,\AA, but where we observe it viewing from the west (right) of the image in 
Figure~\ref{event_05_zu}(e).  In the video accompanying Figure~\ref{event_01_02_zu} 
we have followed the motion of the erupting cool feature with a black arrow, and it tracks
closely in time and horizontal location the base of the spire of the jet feature in 
the XRT video. This spire feature is also apparent in AIA 193 and 211\,\AA\ videos.  
Furthermore, a brightening, consistent with being a JBP brightening, also occurs 
prominently in the XRT videos (and also in the 
AIA videos, with varying degrees of prominence depending on the channel) from very near the location from where
the erupting cool-material feature emanated, and the spire feature moves away from the
potential-JBP brightening.  These characteristics between the erupting cool-material
feature seen here, and the jet-like spire, and the brightenings at the base of the region from
which the cool-material feature erupts, are all common to 
the majority of the cases we have examined of coronal jets resulting from erupting minifilaments 
that produce the jet spires and JBPs at the base location of those erupting minifilaments.  

We have again measured the size and displacement speed of this erupting cool-material feature
for event~1.  From the 304\,\AA\ video, it is changing rapidly from frame to frame over the
period when it is obvious and distinguishable from its surroundings, between 12:12:53 and 
12:15:53\,UT\@.  We measured its size at the single time of 12:14:53\,UT, and found it to be 
about $7''$.  We followed its displacement in 304\,\AA\ images over three different time 
period in the interval 12:12:41 and 12:15:41\,UT, and found a speed of $19.6 \pm 5.3$~\kms.
Although the size is on the small end,
it is still within the ranges obtained in our earlier studies of erupting minifilaments that 
caused X-ray jets (with the values given in \S\ref{subsec-results_e3}), and the speed is
also well within the ranges of those previously measured values for such jets.

Because the quantitative properties of the erupting cool-material feature corresponding to event~1
are consistent with those of erupting minifilaments that produce X-ray jets, and because the 
morphological relationship of that erupting feature to the jet-like spire and base brightening
resembles that of our previously observed jet-producing erupting minifilaments, we again conclude
that the erupting cool-material feature is consistent with being an erupting minifilament that 
produces a coronal jet, and that the source of the EIS event~2 Doppler outflows is this coronal jet.  Thus we support the \cite{schwanitz.et21} conclusion that the probable source of this outflow is a jet, and we have found that it operates in the same manner as other coronal jets that we have observed.

\subsubsection{Event~2}
\label{subsubsec-results_e2}

Figures~\ref{event_01_02_zu}(g---l) show event~2. A white arrow points to a feature that appears to be a jet spire in the XRT image 
in~\ref{event_01_02_zu}(i), and the same feature is visible in~\ref{event_01_02_zu}(l).  
Also in Figure~\ref{event_01_02_zu}(i),
a blue arrow points to a brightening that appears as a JBP at the base of that spire.  
From the animation corresponding to Figure~\ref{event_01_02_zu}, the apparent spire and JBP-like brightening show transient
behavior typical of coronal jets, albeit substantially weaker than jets typically selected for
jet analysis.  In this case the intensity of the spire (faintly appearing over about 13:09 and 13:30\,UT in
the XRT animation, and pointed to by the white arrow in Fig.~\ref{event_01_02_zu}(i)) is too weak to confirm its movement toward, away
from, or stationary relative to, the JBP location.

For this case, a corresponding feature in AIA images is less obvious than we found for the other events.
Careful inspection however of the 171~\AA\ images does show an absorption feature being
ejected, which we indicate with the black arrow in~\ref{event_01_02_zu}(g).  We track this feature with
a black arrow in the animation corresponding to Figure~\ref{event_01_02_zu}.  Its movement in the
animation is closely in sync with the jet spire in the XRT panels.  In addition to 171, this absorbing
feature is also faintly apparent in 193 and 211\,\AA\ animations.  Thus this feature appears to be
a cool-material erupting minifilament that accompanies the jet seen in the XRT images and animation, and
which corresponds to the EIS outflows.  In this case, the erupting minifilament is nearly lost in the
coronal haze in the 171, 193, and 211\,\AA\ images.   

No corresponding feature is discernible from
the background in the 304\,\AA\ images and animations. This is not surprising, as this jet's spire in
X-ray images remains narrow compared to the extent of that jet's base, and hence it would be 
classified as a ``standard jet,'' in the terminology of \citet{moore.et10} and \citet{moore.et13}, and those 
studies found that such standard
jets often lack a significant corresponding jet spire in 304\,\AA\ images.  (See 
\citeauthor{sterling.et22}~\citeyear{sterling.et22} for further examples and further discussions
on standard and blowout jets.)

In this case, it is likely that we are seeing the erupting minifilament at a different stage of its
evolution than the other cases we have examined.  Specifically, it looks as if this eruption is further 
underway by the time we can detect it, compared to other cases where the eruption seemed to be at its 
earliest stages.  Therefore we do not include measurements of its size and speed of this feature for this
event.  Again, however, for this case also, this coronal jet is a candidate for being the source of
the EIS Doppler outflows for event~2.

\section{Summary and Discussion}
\label{sec-discussion}

We have examined five locations in a polar coronal hole that displayed strong transient outflows
of coronal material in \hinode/EIS Doppler images that were found by \citet{schwanitz.et21}. They
investigated 14 events in total, and here we concentrated on the first five events in their list,
which were the ones with corresponding soft X-ray imaging data from \hinode/XRT\@.    For our 
investigation here we primarily used XRT images, along with EUV images from \sdo/AIA, to 
investigate in more detail the nature of the source regions for the outflows. 
\citet{schwanitz.et21} 
attributed the source of these five events to, respectively for events~1---5: ``obvious jets,"
``small-scale eruption," ``small-scale brightening," ``small-scale eruption," and ``unclear."  
Our closer examination here finds all five events to be consistent with being 
coronal X-ray jets.  These jets are smaller-scale, fainter, and harder to detect than those
in studies that specifically aimed at examining obvious jets.  But upon close examination,
the source locations for four of five of these events show quantitative and morphological similarities
to the more-readily observed jets.  The one remaining event, event~2, shows 
morphological similarities with more-readily observed jets, but it was too faint for us to obtain supporting 
quantitative measurements of its corresponding erupting minifilament's size and speed.

For all five events, we found dynamic absorbing features in the EUV images in the source regions 
identified by \citet{schwanitz.et21}.  For
events~1, 3, 4, and 5, these outflowing events appeared extremely similar morphologically to the 
erupting minifilaments that we have observed numerous times at the base of clear coronal jets. 
Measurements of the erupting absorbing feature's length projected against the solar disk and their
outward-moving speed values fell within the range that we had measured for the erupting minifilaments
causing X-ray jets in polar coronal holes in our previous studies \citep{sterling.et15,sterling.et22}.
Event~2 shows a similar erupting feature being expelled from the source region, although somewhat harder
to see (likely due to obscuration by foreground material) and perhaps later in its eruption than 
for the other events.
For each of these five events, further inspection of the soft X-ray XRT images and animations showed
a spire-like feature, and a JBP-like brightening near the spire's base.  Moreover, this spire 
moved away from the JBP in the four cases where the spire was bright enough to make a determination 
(all except for event~2).  In addition, dimming signatures in events~3, 4, and~5, are consistent 
with material from the low corona being expelled in the jet episodes.  (Candidates for similar 
dimmings exist in the events~1 and~2 XRT videos as well.)  These morphological and measured 
characteristics are all consistent with the behavior of jets that fit the minifilament-eruption picture.

A common characteristic of jet-producing erupting minifilaments observed in on-disk jet events is
that they originate from neutral lines between opposite-polarity flux patches, and furthermore,
magnetic flux cancelation often occurs along that neutral line leading up to and during the time of 
the minifilament-eruption 
onset \citep[e.g.,][]{shen.et12,adams.et14,panesar.et16a,panesar.et17,panesar.et18a,sterling.et16b,sterling.et17,mcglasson.et19,muglach21}. 
Although all of our events occur in a north polar coronal hole, one of these, event~5, is far
enough south and has strong enough fields at its base for us to make some inferences about the
erupting minifilament's magnetic environment. We find that the minifilament of the 
region indeed erupted within a few arcseconds of the neutral line of a bipolar pair
of opposite-polarity magnetic patches.  Moreover, in the hours prior to the minifilament's
eruption and jet production, that pair of patches approached each other, consistent with 
flux cancelation being
the cause of the minifilament's eruption.  Thus these magnetic field observations are also
consistent with the event~5 EIS outflow source region being a coronal jet.

One might ask: If the source of the strong EIS Doppler outflowing events are typical coronal jets,
then why are those jets not more obvious as typical jets?  In other words, why do we have to look
at them with the deep analysis presented here before we can recognize them as coronal jets?  There
are two points to address in answering this: one is the nature of the typical jets of 
other studies compared with the jets identified in this study, and a second is why EIS happened 
to find the specific outflows examined in the \citet{schwanitz.et21} study.

Regarding the first point: A key difference between the jets found here and those found in other studies is that
the other studies all started out specifically looking for jets.  Thus, they examined jets that were
comparatively easy to identify in images.  The X-ray jets of \citet{sterling.et15} were ones that were well observed
by \citet{moore.et13} in XRT images.  Similarly, several other studies 
\citep{sterling.et16b,sterling.et17,sterling.et22,baikie.et22}
selected jets specifically because they were relatively well seen in XRT images.  (Actually, \citeauthor{sterling.et22}~\citeyear{sterling.et22}
deliberately included a few jets that they described as ``less prominent yet still distinct in XRT" in their
study.  Nonetheless those jets, which included J7, J8, J12, and J19 in their study, were still 
easier to identify in XRT images than the five jets of the present study.)  All of those studies selected
their jets by starting with movies formed from XRT images and specifically looking for jets.
Similarly, there have been many studies of jets that began by looking for a set of jets in EUV movies
\citep{nistico.et09,panesar.et16a,mulay.et16,panesar.et17,panesar.et18a,mcglasson.et19,kumar.et19},
and many additional studies looking at only about one or two jets, most of which began by 
looking for well-observed jets.
Here, our approach was different, in that we did not specifically look for jets to study.  Instead, 
interesting features with strong-outflow Doppler signatures in EIS 
spectral scans were first identified in the EIS observations, and then the solar coronal 
source for those regions were searched for in the EUV and SXR imaging data.  What we found 
were disturbances that have the properties of coronal jets in the source regions, but
those jet characteristics happen to be much harder to detect than in studies of jets found by 
specifically searching images for obvious jets.

Now on to the second point: The reason \citet{schwanitz.et21} only find outflows from
jets that were comparatively weak and hard to detect is because EIS has a limited FOV and a 
low time cadence.  The jet occurrence rate of about 30/day in a coronal hole deduced by
\citet{savcheva.et07} is for jets that are relatively easy to see.  From this we can expect
about one such easy-to-see jet per hour in the entire coronal hole.  Each EIS scan covers only $322'' \times 384''$, according to the information in the EIS study used for the observations (which was ``{\it HOP81\_new\_study} with number 582," from \citeauthor{schwanitz.et21}~\citeyear{schwanitz.et21}), over about 70~minutes.  The coronal hole studied by
\citet{savcheva.et07} \citep[see][]{cirtain.et07} had about twice that extent in the east-west 
direction and so there is only a small chance that the $2''$-wide rastering EIS slit would pass
over a substantial jet occurring about once an hour at a random location over the entire coronal hole and
lasting only a few minutes.  Thus EIS (operating in this mode) could easily miss a jet that would have
been significant enough in X-ray images to be logged by studies specifically looking for such jets.
Apparently instead, however, EIS found outflows from jets that were much less obvious (and likely more 
frequent) than those usually
found in studies deliberately targeting X-ray jets.

These findings have significant implications for the occurrence frequency for outflow-producing
coronal jets.  The occurrence rates for jets of  \citet{savcheva.et07}, of $\sim$60/day in two 
polar coronal holes,  relies
upon jets that appeared as ``visible ejection of material on a timescale of several  tens of
minutes" and that showed ``a rapid increase in the length of [a jet-spire-like] brightness
enhancement'' with  time.  Thus, the jets they selected were, as would be expected from the 
above discourse, those
that appeared clearly to be jets in XRT images.  We have shown here that there are many 
features that are difficult to see in XRT images, but which are driven by the same processes
that make the easier-to-see jets.  Another factor is that several of the features we found 
were shrouded in quiet Sun ``coronal haze."  This factor alone could be largely responsible
for the difficulty in identifying the jets.  

Jets that appear in coronal holes unobscured by such haze are easier to see; this ease of
visibility in coronal holes compared to quiet Sun was pointed out by \citet{moore.et13}, and also
discussed in \citet{baikie.et22}.  In addition  to this being a consideration in the jet-count
statistics of \citet{savcheva.et07}, it also  could explain why they found most of their jets to
occur in polar coronal holes; even if jets exist in equal numbers in quiet Sun outside of those coronal hole
regions, they may be harder to detect.  A similar phenomenon was found with X-ray bright points
(XBPs), which are X-ray-bright compact features mainly seen in polar coronal holes. 
\citet{nakakubo.et00} found that XBP occurrence frequency seemingly is inversely related to the
number of sunspots over a solar cycle, but they  concluded that the apparent increase in XBPs
during low-spot times was likely due to a decrease in the X-ray background intensity, rather than
due to an actual increase in XBPs.   Therefore, it is plausible that many coronal jets exist but
are difficult to detect due to  obscuration by foreground quiet-Sun coronal haze.  

\begin{figure}
\hspace*{3.0cm}\includegraphics[angle=0,scale=0.75]{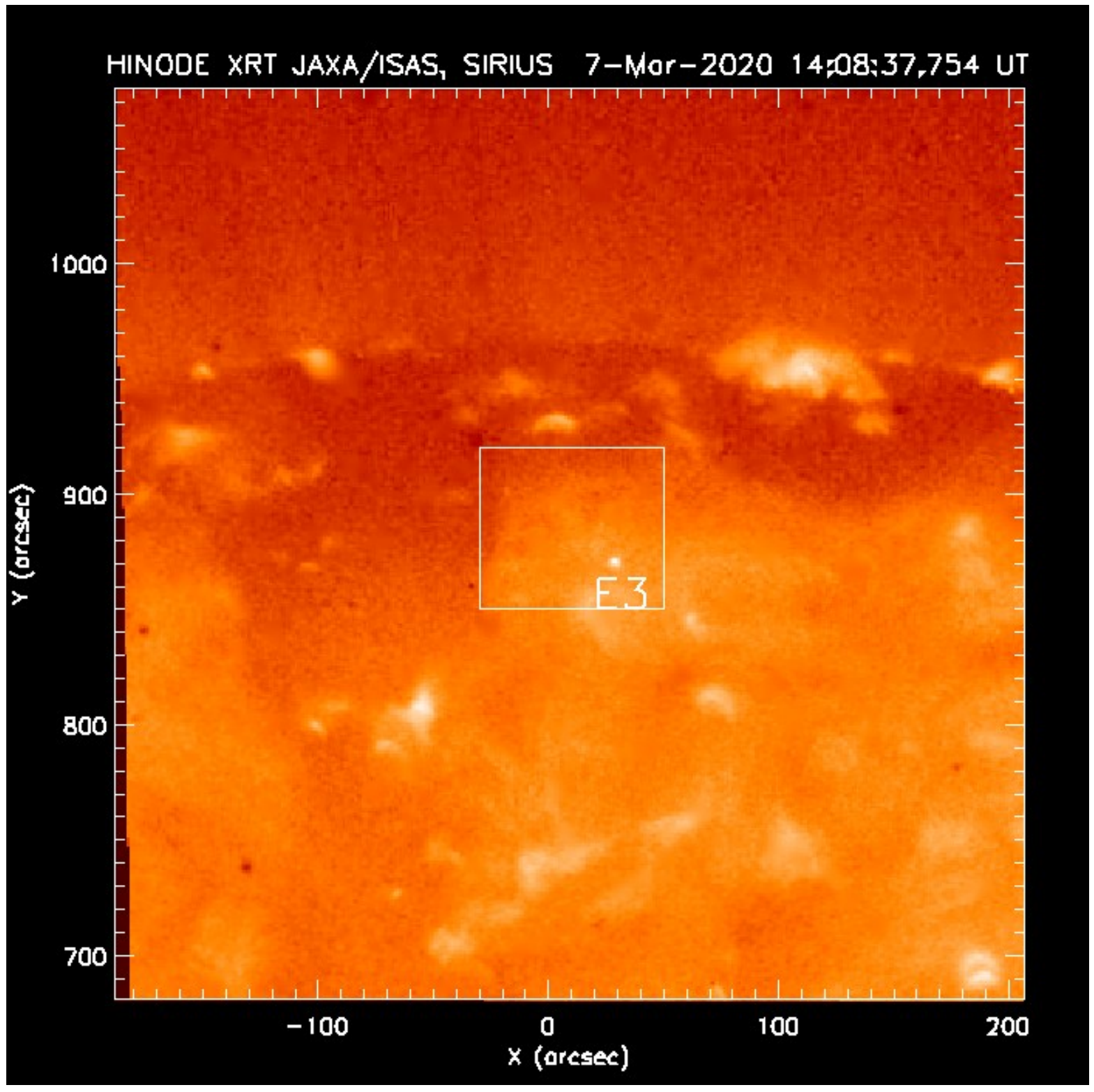}
\caption{
A soft X-ray image from the \hinode\ XRT instrument's Al~Poly filter, of the north polar coronal 
hole region on 2020 March~7, showing an overview of the northern polar coronal hole region where all of the events studied here occurred.  One 
of the events of Table~1, labeled E3, is in progress, and the white box shows the approximate field of view displayed for the corresponding EIS-outflow event in Fig.~\ref{eis_overview}.   North is up and west is to the right 
in this and in all 
other solar images and animations of this paper.  Dark spots in the image, e.g.\ at (-140,740), 
and similar patches and blotches, are artifacts. The accompanying animation shows the XRT movie for
this period, covering 2020 March~7 over 12:00:01---18:45:20\,UT, but with a gap between 14:34:46 
and 16:30:15\,UT because there are no events in our data set over that interval.  That animation contains labels and boxes similar to those shown for event~3 in this figure, but for all five events, with the coordinates of the
boxes in Fig.~\ref{eis_overview} giving the approximate locations of those boxes for each respective jet in the movie.
In that animation at the time of event~5, the ``E5" label is located near a brightening that is the suspected source region of the outflows, which is located about 30$''$ south of the EIS box showing the
EIS Fig.~\ref{eis_overview} panel's FOV; see \S\ref{subsec-results_e5} for discussion of the spire and observed EIS outflows being separated from the brightening.  The 
entire movie runs for 26\,s.
}
\label{xrt_overview_zu}  
\end{figure}

\begin{figure}
\hspace*{3.0cm}\includegraphics[angle=0,scale=1.2]{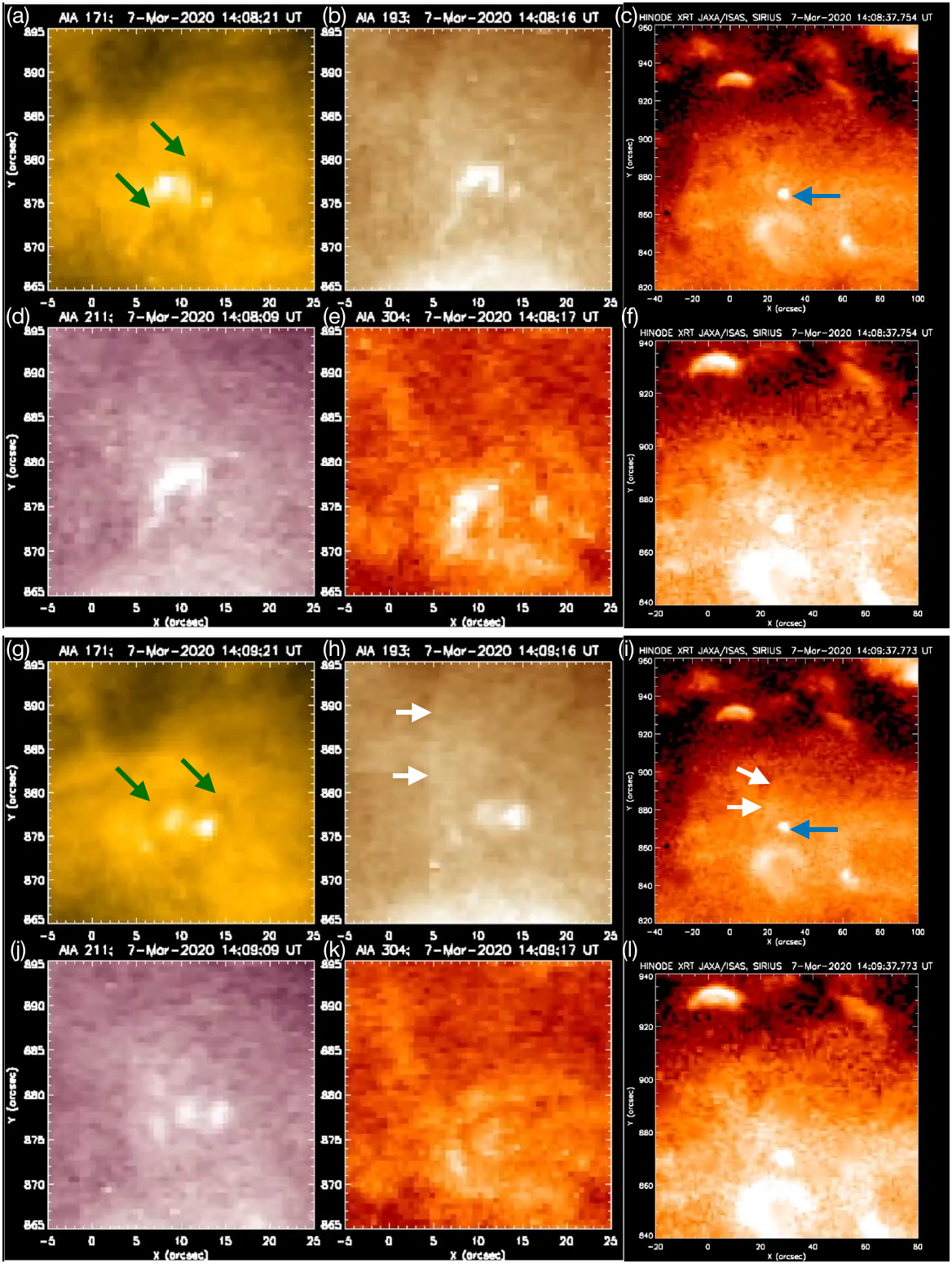}
\caption{
Zoomed-in views of Table~1 event~3, at AIA and XRT wavelengths.  Panels~(a)---(f) show the region
at about the same time, in (a) AIA~171\,\AA, (b) AIA~193\,\AA, (c) XRT ``thin~Al," (d) AIA~211, 
(e) AIA~304, and (f) XRT thin~Al images.  The field of view (FOV) is the same for all of the AIA
images.  For XRT the FOV is larger than that of AIA; both XRT images are at the same time, with 
panel~(f) zoomed-in more than~(c), and (f) also uses an intensity scaling set to enhance 
fainter aspects of the feature.    Panels~(g)---(l) mimic the ordering and parameters of the images
of (a)---(f), but with all of those panels at nearly the same time (within a few seconds of each 
other), at a later time (about one minute later) than the first six panels.  Green arrows show 
a cool-material erupting minifilament, at two different times in~(a) and~(g).  Blue arrows in~(c) 
and~(i) show brightening similar to a coronal-jet jet bright point (JBP)\@.  White arrows in 
(h) and (i) point to a spire that moves
away from the JBP location, indicating that the event is likely a faint coronal jet.  All of these panels (a---l) are from near the time of the center of the EIS blueshifted contour in Fig.~\ref{eis_overview}, which was at about 14:09\,UT (table~\ref{tab:table1}).   The accompanying 
animation shows the evolution of these panels, covering 2020 March~7 over 13:49---14:35\,UT\@.  The 
entire movie runs for 3\,s.
}
\label{event_03_zu}  
\end{figure}

\begin{figure}
\hspace*{1.0cm}\includegraphics[angle=0,scale=0.6]{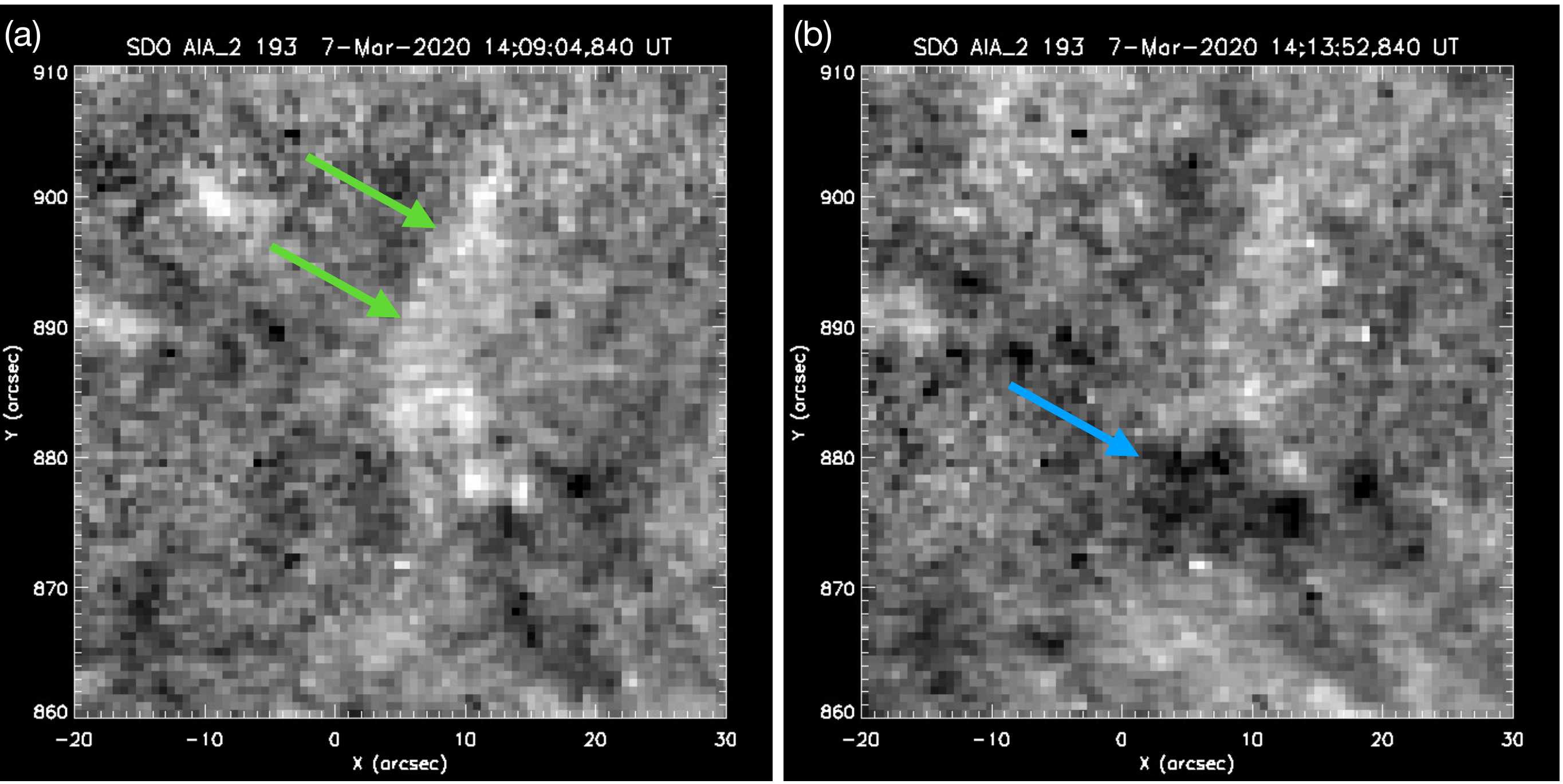}
\caption{
AIA 193~\AA\ difference images of Table~1 event~3.  These are fixed differences, whereby 
we have subtracted the image at 13:49:04\,UT in both frames.  Panel~(a) shows what appears to
be a spire of a typical coronal jet (green arrows).  Panel~(b) shows a dimming region at 
the base of the apparent spire (blue arrow). The accompanying animation shows the time 
evolution of difference images, with the same initial frame used here subtracted.  The 
movie covers covering 2020 March~7 over 13:49---14:22\,UT, and runs for 3\,s.  
}
\label{event_03_dim_zu}  
\end{figure}

\begin{figure}
\hspace*{3.0cm}\includegraphics[angle=0,scale=1.2]{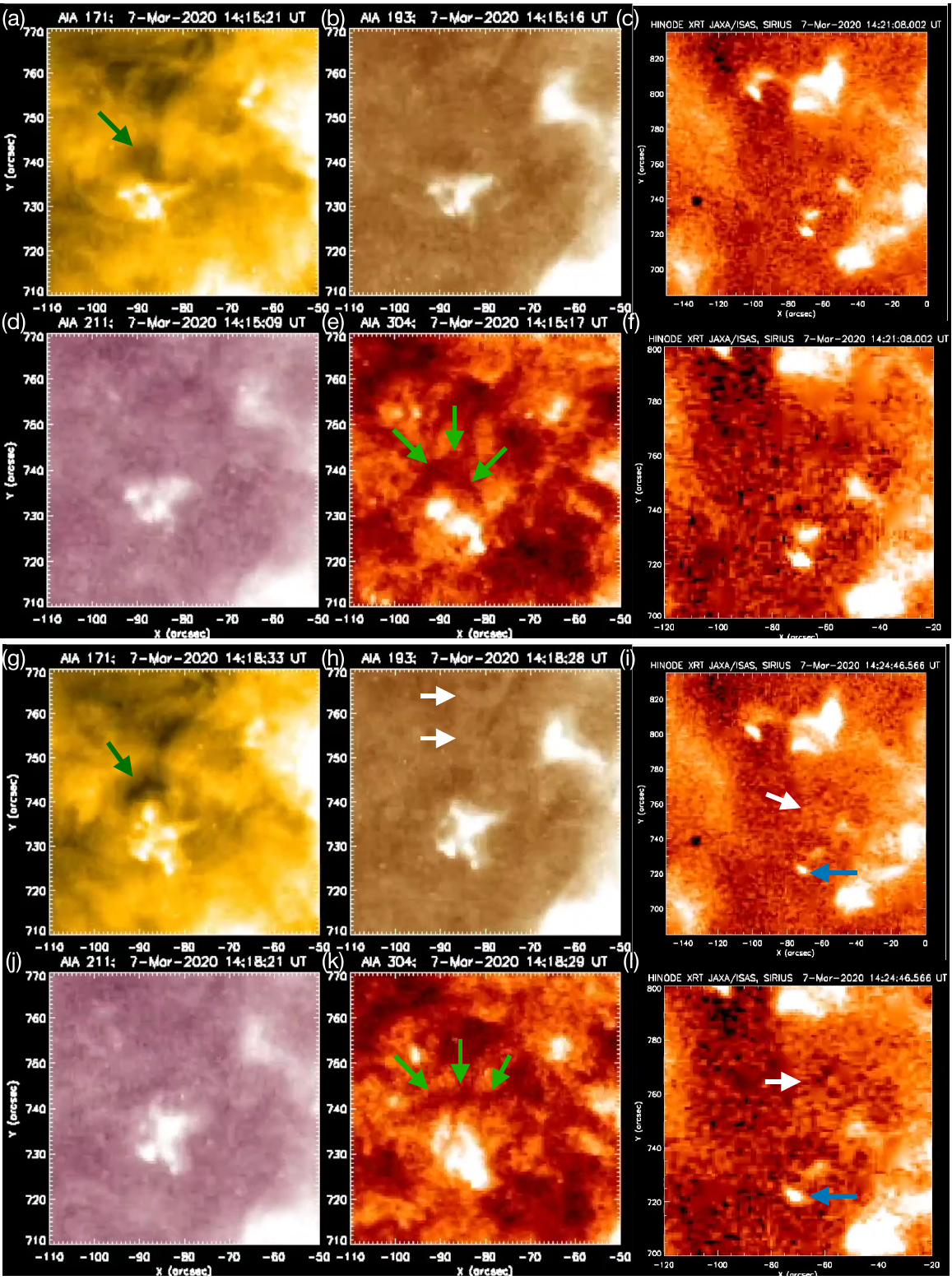}
\caption{
Zoomed-in views of Table~1 event~4, at AIA and XRT wavelengths.   The basic layout of
the panels is as described
in Figure~\ref{event_03_zu}.  Green arrows show 
a cool-material erupting minifilament, at about the same time in~(a) and~(e), and at a later
time in~(g) and~(k).  Blue arrows in~(i) 
and~(l) show a brightening similar to a JBP\@.  White arrows in (h) and (i) point to a faint feature
that resembles a coronal-jet spire.  The accompanying 
animation shows the evolution of these panels, covering 2020 March~7 over 14:00---14:48\,UT\@.  The 
entire movie runs for 3\,s.   From table~\ref{tab:table1} the center of the blueshift contour is at about 14:29\,UT, at which time the movies still show apparent outflow. 
}
\label{event_04_zu}  
\end{figure}

\begin{figure}
\hspace*{3.0cm}\includegraphics[angle=0,scale=1.2]{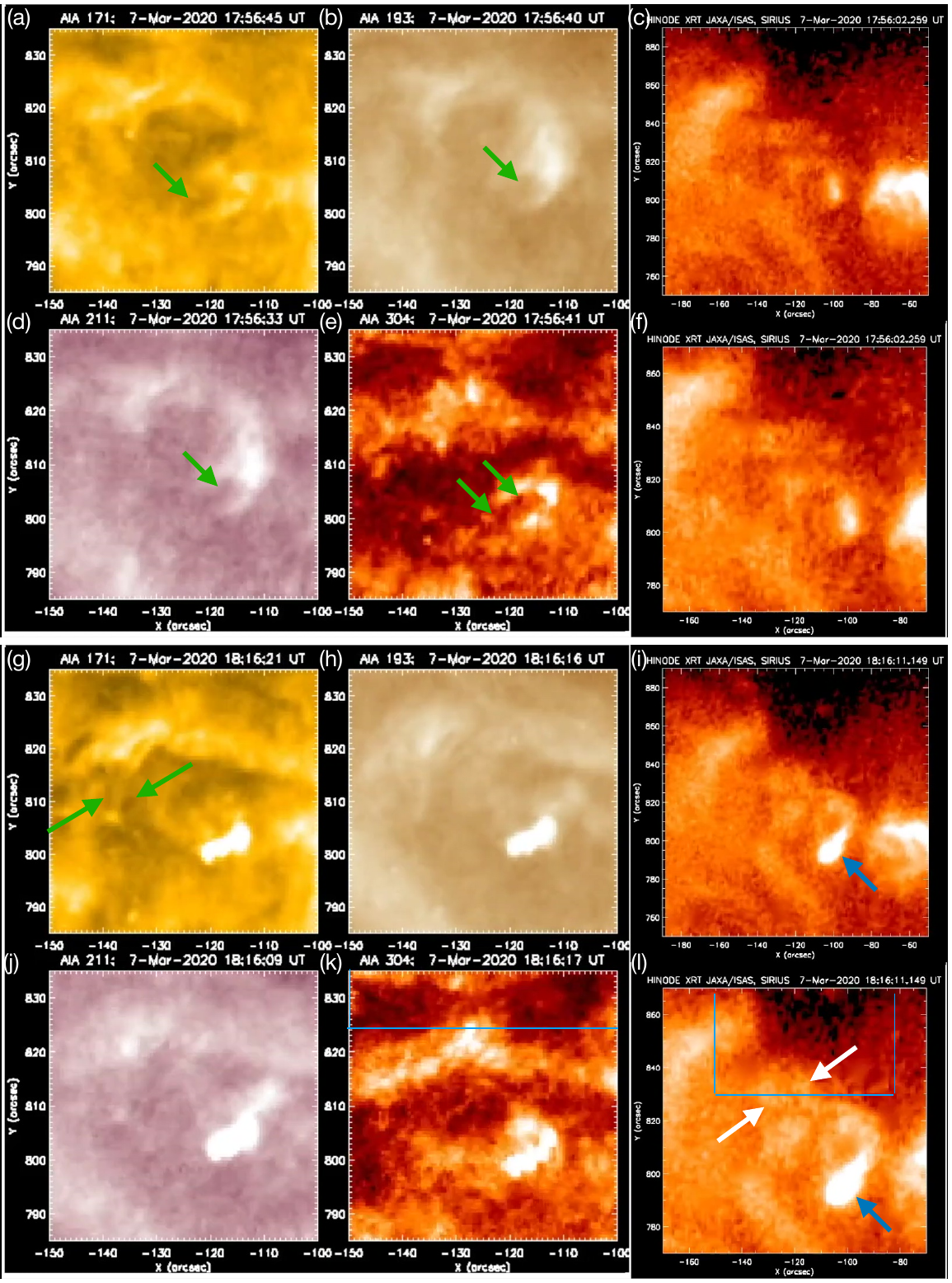}
\caption{
Zoomed-in views of Table~1 event~5, at AIA and XRT wavelengths.  The basic layout of
the panels is as described 
in Figure~\ref{event_03_zu}.  Green arrows show 
a cool-material erupting minifilament, at about the same time in~(a), (b), (d), and~(e).  
Blue arrows in~(i) 
and~(l) show a brightening similar to a JBP, and that brightening is visible but saturated in 
the AIA images in (g), (h), (j), and~(k).   White arrows in (l) point to either side of 
a faint feature that resembles a coronal-jet spire.  Blue lines in (k) and (l) represent portions of the EIS FOV shown in Fig.~\ref{eis_overview}, and which is shown as a white box in the video accompanying Fig.~\ref{xrt_overview_zu}. The  
animation accompanying the current figure shows the evolution of these panels, covering 2020 March~7 over 17:47---18:44\,UT\@. The animation runs twice, where the first time through an arrow in the 171~\AA\ panel shows outflows from the minifilament-eruption location that reach out to the location near where the spire forms, and the blue lines in the XRT panel represent the FOV of the panel for event~5 in Fig.~\ref{eis_overview}.  The second time through the animation plays without the arrow and box overlays. The 
entire movie runs for 6\,s.   From table~\ref{tab:table1} the center of the blueshift contour is at about 18:32\,UT, at which time the XRT movie is still consistent with continued outflow. 
}
\label{event_05_zu}  
\end{figure}

\begin{figure}
\hspace*{2.0cm}\includegraphics[angle=0,scale=1.0]{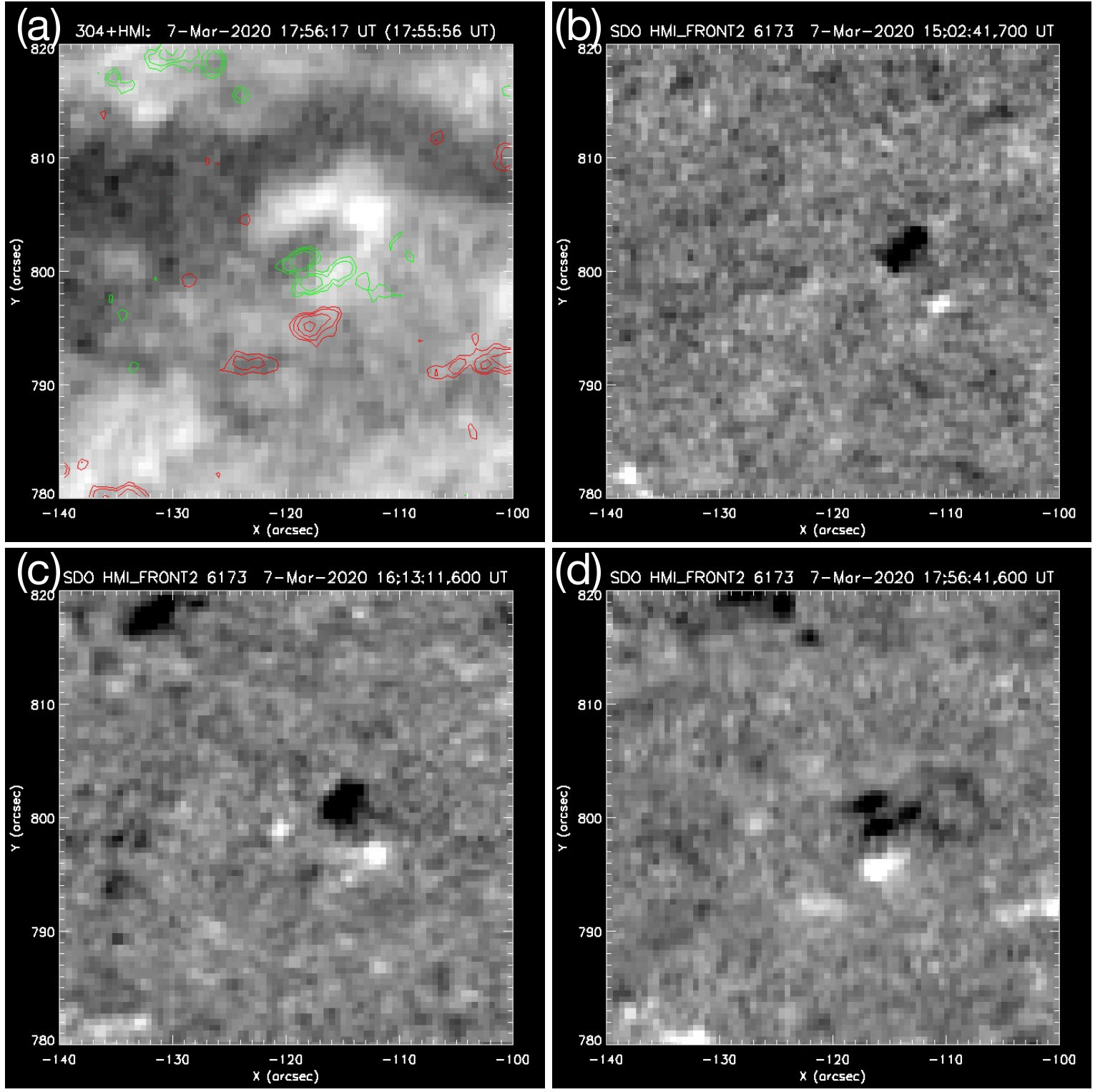}
\caption{
Magnetic field evolution at the base of event~5. Panel~(a) shows positive (red) and negative (green)
contours from an HMI line-of-sight magnetogram at 17:55:56\,UT on 2020 March~2, as contours 
with levels set at 30 and 50\,G\@.  This is overlaid onto a grayscale version of an 
304\,\AA\ image within seconds of that in Fig.~\ref{event_05_zu}(e) (the FOV here is more zoomed in than shown in Fig.~\ref{event_05_zu}(e)), just as the minifilament is erupting.  That erupting
minifilament plausibly formed and was ejected from the neutral line of the two closest opposite-polarity
strong flux patches.  Panels~(b)---(d) show the line-of-sight HMI magnetograms of the location evolving over time, with positive and negative polarities represented by white and black, respectively (contours in 
(a) are within a minute of (d)).  These panels show that the two opposite-polarity patches that may have caused the minifilament eruption converged toward each other in the hours prior to that eruption.  Such flux convergence (and cancelation) is commonly observed to occur leading up to minifilament eruptions 
that produce typical coronal jets. The FOV here is slightly smaller than that of the AIA images in Fig.~\ref{event_05_zu}, in order to center the
bipolar flux patches better.
}
\label{event_05_hmi_zu}  
\end{figure}

\begin{figure}
\hspace*{3.0cm}\includegraphics[angle=0,scale=1.2]{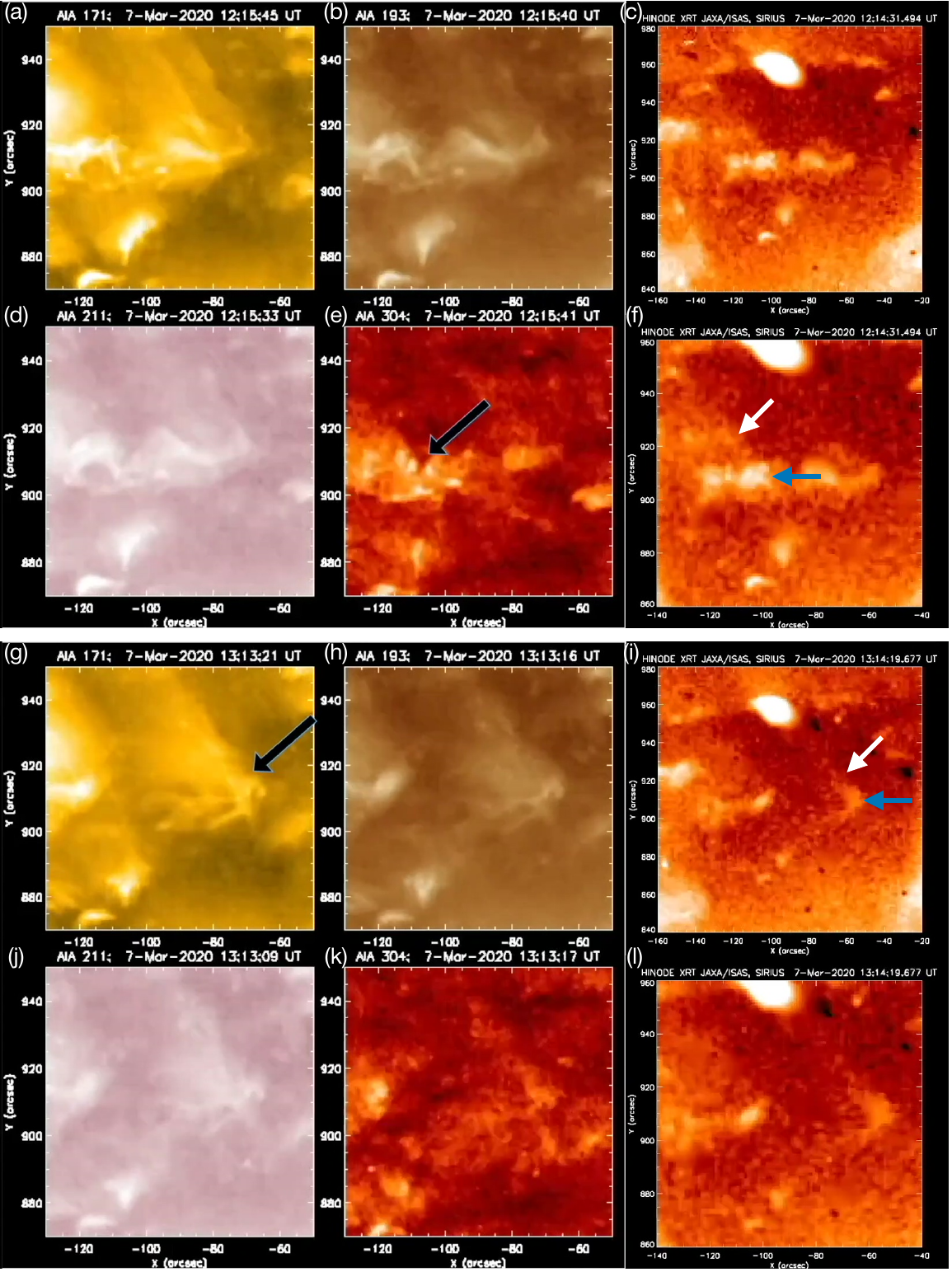}
\caption{
Zoomed-in views of Table~1 event~1 (a---f) and event~2 (g---l), at AIA and XRT 
wavelengths.  The basic layout of the panels is as described in
Figure~\ref{event_03_zu}.  The black arrow in~(e) shows  a cool-material feature in an AIA
304\,\AA\ image that is consistent with being an erupting minifilament,  producing the
weak jet-spire-like feature pointed to by the white arrow in the XRT frame in panel~(f), 
which feature coincides
in time and location with the outflows observed in EIS for event~1.  The blue arrow in (f)
points to a brightening similar to a JBP adjacent to the spire-like feature.  The black
arrow in~(g) shows a cool-material  feature in an AIA 171\,\AA\ image that is consistent
with being an erupting minifilament,  producing the weak jet-spire-like features pointed
to in the XRT frame in panel~(i), and which coincides in time and location of the outflows
observed in EIS for event~2. The blue arrow in (i) points to a brightening similar to a
JBP adjacent to the spire-like feature.  The accompanying  animation shows the evolution 
of these panels, covering 2020 March~7 over 12:00---13:57\,UT\@.  The animation plays
twice: the first time through the black arrows of panels~(e) and~(g) are overlaid, and
trace the movement of the cool-material features in time; the second time through the 
animation plays without any overlays.  The  entire movie 
runs for 8\,s. From table~\ref{tab:table1} the center of the blueshift contour for events~1 and~2 are respectively at about 12:18 and 13:19\,UT, and the movie shows features consistent with outflows from the respective locations at those times. 
}
\label{event_01_02_zu}  
\end{figure}

In addition,
it is likely that there are many small-scale coronal jets that are difficult to detect in
X-rays because they are intrinsically faint.  Both of these factors could contribute to an
underdetection of the true frequency of coronal jets in X-ray images, and also in EUV images.  Revisiting
the video accompanying Figure~\ref{xrt_overview_zu}, it is apparent that only events~1 and~2 
occur in a coronal-hole-appearing region.  All of the other events either occur in quiet Sun, or
perhaps in the polar coronal hole but with quiet-Sun corona in the foreground that partially
obscures the event.  This may explain why only event~1 was described as 
being due to ``obvious jets" by \citet{schwanitz.et21}.  Event~2 also occurs in a coronal-hole-appearing
location, but it was perhaps intrinsically too faint for a clear identification with a specific jet without 
the closer analysis performed here.

In addition to coronal jets observed in XRT images, there are smaller-scale jet-like features,
called ``jetlets" that appear at the base of polar plumes \citep{raouafi.et14}.  
More generally, jetlets occur on the edges of the chromospheric network, and they may operate 
in the same manner as coronal jets \citep{panesar.et18b,panesar.et19}.  Perhaps-similar
features  appear in active-region plage \citep{sterling.et20c}, and in ultra-high-resolution
EUV images of features dubbed ``campfires" \citep{panesar.et21}.  On still smaller scales,
there is the possibility that some fraction of spicules are produced in the same manner as
coronal jets and jetlets, viz., via the minifilament-eruption mechanism operating on a much smaller
size scale, via eruption of ``microfilaments" \citep{sterling.et16a,sterling.et20b}.

If all of these jet-like features, including coronal jets, jetlets, and some fraction of 
spicules, operate in the same fashion as coronal jets, then it is possible that they all produce 
outflows similar to those detected from
EIS, which here we argue are weak X-ray coronal jets.  Then, in aggregate, these jet-like 
features resulting from small-scale eruptions might contribute substantially to the solar wind
outflow.  They may also contribute to the population of switchbacks that are observed in the
solar wind  observed by PSP \citep{horbury.et20,sterling.et20b,neugebauer.et21}, and even
 much earlier by \citet{neugebauer12} based on observations of switchback-like phenomena in Ulysses 
 data.  Previously it had been suspected that coronal jets were not frequent enough to
account for the large quantity of switchbacks observed by PSP\@.  But it may be that the
assumed number of jets is severely underestimated.  To address the question directly will 
require development of methods to estimate the true rate of jet production.  This requires
estimating the effect of coronal haze on the counts, and further confirmation that the 
smaller-scale features, such as jetlets and spicules, are in fact due to eruption of small-scale
filament-like features, triggered to erupt by magnetic flux cancelation at their bases.  

It is also desirable to observe spectroscopically, with spectral scans such as those
done with EIS by \citet{schwanitz.et21}, a bone fide ``typical" coronal jet.  It would be of
interest to see whether such a jet results in outflows similar to those of the  
\citet{schwanitz.et21} events, except perhaps stronger (which might be evidenced by a 
stronger emission-measure depletion, for example).
Such observations could provide additional support that \citet{schwanitz.et21} features, at 
least the five examined here, are indeed small-scale version of such bona fide coronal 
jets.  It will also be of interest to see how the outflows here connect to other outflow events
commonly observed with EIS and similar spectrometers \citep[e.g.][]{harra.et08,brooks.et11,harra.et12,brooks.et20,tian.et21,yardley.et21}.  It is also
of interest to investigate whether the EIS Doppler features referred to as dark jets by \citet{young15}
might be similar to the features we observe here.

\begin{acknowledgments}
A.C.S., R.L.M. and N.K.P. received funding from the Heliophysics Division of NASA's Science 
Mission Directorate through the Heliophysics Supporting Research (HSR, grant No.~20-HSR20\_2-0124) Program, 
and the Heliophysics Guest Investigators program.  A.C.S. and R.L.M. also received support from
the Heliophysics System Observatory Connect (HSOC, grant No.~80NSSC20K1285) 
Program.  A.C.S. received additional support through
the MSFC \hinode\ Project, and N.K.P. received additional support through a NASA \sdo/AIA grant.  
\hinode\ is a Japanese mission developed and launched 
by ISAS/JAXA, with NAOJ as domestic partner and NASA and UKSA as international 
partners. It is operated by these agencies in co-operation with ESA and NSC (Norway). 
We acknowledge the use of AIA data. AIA is an instrument onboard \sdo, a mission of
NASA's Living With a Star program.
\end{acknowledgments}

\bibliography{ms}

\clearpage

\begin{deluxetable}{lllllll@{\hskip 1.5cm}l}
\tabletypesize{\scriptsize}
\tablecaption{EIS Outflow Events Observed with XRT and AIA \label{tab:table1}}
\tablehead{
\colhead{Event}& \colhead{EIS Time (UT)\tablenotemark{a}}& \colhead{EIS Size (arcsec$^2$)\tablenotemark{a}} & \colhead{XRT
Time (UT)\tablenotemark{b}}&  \colhead{XRT Location\tablenotemark{b}} &  \colhead{EMF Size (km)\tablenotemark{c}}&  \colhead{EMF Velocity (\kms)\tablenotemark{c}}&\colhead{EMF Notes\tablenotemark{d}}
}
\startdata
1 & 12:17:55 &   3244 & 12:14---12:29 & (-100,900)& $\approx$5000 & $19.6 \pm 5.3$ &\parbox{3cm}{\vspace{0.3cm}Small-sized EMF visible in 304\,\AA\@.\vspace{0.3cm}} \\
2 & 13:18:39 &	\ 677 &	13:09---13:22 & (-60,900) &   Uncertain	 & Uncertain &\parbox{3cm}{\vspace{0.3cm}Strands of an EMF visible in 171\,\AA\@.\vspace{0.3cm}} \\
3 & 14:09:00 &	\ 245 &	14:00 ---14:14 	& (30,860) &  $5700\pm 690$   & $46\pm 15 $ &\parbox{3cm}{\vspace{0.3cm}EMF visible in all channels. Confined eruption.\vspace{0.3cm}} \\
4 & 14:28:34 &	\ 721 &	14:14---14:32 	& (-70,720) &  $12{,}600 \pm 110$ & $29\pm 10$ &  \parbox{3cm}{\vspace{0.3cm} Clear EMF in all channels, maybe best in 304\,\AA\@.\vspace{0.3cm}}\\
5 & 18:32:16 &	\ 365 &	18:00---18:45+ & (-199,800) &  $14{,}000\pm 420$	 & $32 \pm 11$ & \parbox{3cm}{\vspace{0.3cm}Clear EMF in all channels, maybe best in 304\,\AA\@. Appears to be ejective.\vspace{0.3cm}} \\
\hline
\hline
\enddata
\tablenotetext{a}{From Table~2 of \citet{schwanitz.et21}.}
\tablenotetext{b}{XRT jet times and locations, in arcseconds from disk center.  Locations are approximate, based on video in Fig.~\ref{xrt_overview_zu} 
(which is rotated to 2020 March~7, 15:00~UT\@).}
\tablenotetext{c}{Size and velocity of erupting minifilament (EMF) from AIA images, with 1$\sigma$ standard deviations where determined (see text for details).}
\tablenotetext{d}{Notes on detectability of EMF in AIA images.}
\end{deluxetable}
\clearpage

\comment{

\begin{deluxetable}{lll@{\hskip 1.5cm}l}
\tabletypesize{\footnotesize}
\tablecaption{EIS Outflow Events Observed with XRT and AIA \label{tab:table2}}
\tablehead{
\colhead{Event}& \colhead{EMF Length (km)} &\colhead{EMF Velocity (\kms)}&\colhead{Notes}
}
\startdata
1 &  $5700\pm 690$   & $46\pm 15 $	& Uncertain \\
2 &  $12,600 \pm 110$ & $29\pm 10$	 & Partial \\
3 &  $14,000\pm 420$	 & $32 \pm 11$	& Confined \\
4 &  $\gtsim 5000$ & $19.6 \pm 5.3$	&  \\
5 &   Uncertain	 & Uncertain	& Ejective \\
\hline
\hline
\enddata
\end{deluxetable}
\clearpage

\begin{deluxetable}{llllll@{\hskip 1.5cm}l}
\tabletypesize{\footnotesize}
\tablecaption{EIS Outflow Events Observed with XRT and AIA \label{tab:table1}}
\tablehead{
\colhead{Event}& \colhead{EIS Blueshift Time (UT)}& \colhead{EIS Size (arcsec$^2$)} & \colhead{XRT
Event Time (UT)} & \colhead{EMF Length (km)} &\colhead{EMF Velocity (\kms)}&\colhead{Notes}
}
\startdata
1 & 12:17:55 &   3244 & 12:15---12:29 &	 $5700\pm 690$   & $46\pm 15 $	& Uncertain \\
2 & 13:18:39 &	\ 677 &	13:09---13:22(?)& $12{,}600 \pm 110$ & $29\pm 10$	 & Partial \\
3 & 14:09:00 &	\ 245 &	14:00 ---14:14 & $14{,}000\pm 420$	 & $32 \pm 11$	& Confined \\
4 & 14:28:34 &	\ 721 &	14:14---14:32 &	 $\gtsim 5000$ & $19.6 \pm 5.3$	&  \\
5 & 18:32:16 &	\ 365 &	18:00---18:45+ & Uncertain	 & Uncertain	& Ejective \\
\hline
\hline
\enddata
\end{deluxetable}
\clearpage

} 

\clearpage

\clearpage

\clearpage

\clearpage

\clearpage

\clearpage

\clearpage

\clearpage

\end{document}